\documentclass[superscriptaddress,showpacs,showkeys,aps,preprint,longbibliography]{revtex4-1}
\usepackage{graphicx}
\usepackage{dcolumn}
\pdfoutput=1
\newcommand{\angst}{\mathring{\text A}}
\newcommand{\tract}{\!\! -  \!\!}
\newlength{\figwidth}
\begin{document}
\title{Methylammonium fragmentation in amines as source of localized trap levels and 
the healing role of Cl in Hybrid Lead-Iodide Perovskites}
\affiliation{CompuNet, Istituto Italiano di Tecnologia, Via Morego 30, 16163 Genova, Italy}
\affiliation{Istituto Officina dei Materiali, CNR-IOM Cagliari SLACS, 
             Cittadella Universitaria, I-09042 Monserrato  (CA), Italy}
\author{Pietro Delugas}
\affiliation{CompuNet, Istituto Italiano di Tecnologia, Via Morego 30, 16163 Genova, Italy}
\affiliation{Istituto Officina dei Materiali, CNR-IOM Cagliari SLACS, 
             Cittadella Universitaria, I-09042 Monserrato  (CA), Italy}
\author{Alessio Filippetti}
\email[Corresponding author:\ ]{alessio.filippetti@dsf.unica.it}
\affiliation{Istituto Officina dei Materiali, CNR-IOM Cagliari SLACS, 
             Cittadella Universitaria, I-09042 Monserrato  (CA), Italy}
\author{Alessandro Mattoni}
\email[Corresponding author:\ ]{Alessandro.Mattoni@iom.cnr.it}
\affiliation{Istituto Officina dei Materiali, CNR-IOM Cagliari SLACS, 
             Cittadella Universitaria, I-09042 Monserrato  (CA), Italy}

\begin{abstract}
The resilience to deep traps and localized defect formation is one of the important aspects
that qualify a material as suited photo-absorber in solar cell devices. Here we investigate
by ab-initio calculations the fundamental physics and chemistry of a number of possible
localized defects in hybrid methylammonium lead-iodide perovskites. Our analysis
encompasses a number of possible molecular fragments deriving from the dissociation of
methylammonium. In particular, we found that in stoichiometric conditions both ammonia and
methylamine molecules present lone-pair localized levels well within the perovskite band
gap, while the radical cation CH$_2$NH$_3$$^+$ observed by EPR after irradiation injects
partially-occupied levels into the band gap but only in $p$-type conditions. These defects
are thus potentially capable to significantly alter absorption and recombination
properties. Amazingly, we found that additional interstitial Cl is capable to remove 
these localized states from the band gap. These results are consistent with the
observed improvement of photoabsorption properties due to the Cl inclusion in the solution
processing.
\end{abstract}
\maketitle
\section{Introduction}
The outstanding photoconversion properties of $\mathrm{CH_3NH_3PbI_3}$  perovskites
(we will, from now on, abbreviate methyylammonium as M and methylammonium lead iodide as $\mathrm{MPbI_3}$) 
have been subject of an impressive number of experimental and theoretical studies in the last few
years~\cite{Lee2012,Stranks2013,Xing2013,Burschka2013,Xing2014,Mosconi2013}. 
Nowadays, most of the key ingredients which characterize the
photo-absorption properties of the system, at least in its bulk form, are understood: the
large absorption coefficient in the important frequency range~\cite{Filippetti2014,Filippetti2014a,
Yin2014b,Yin2014,DeWolf2014} the dominance
of unbound electron-hole recombination over excitonic effects at room temperature~\cite{
 Saba,D'Innocenzo2014},
the small band-to-band radiative recombination rate~\cite{Filippetti2014a,Tvingstedt2014}, 
the long ($\approx 10^2\div10^3$~ns)  lifetime and diffusion length (hundreds nm)~\cite{
Stranks2013,D'Innocenzo2014,Marchioro2014,Roiati2014,Saba}. 
In summary, the great appeal of this system stems from the fact that, while processed
through low-cost solid solution as organic materials, it is nevertheless assignable, for
what concerns photo-absorption properties, to the category of the best crystalline
semiconducting absorbers (see Refs~\onlinecite{Filippetti2014a,Filippetti2014,Yin2014b}
for a detailed comparison with GaAs).

However, several aspects are still to be fully understood. One of them concerns the
incidence and typology of native defects present in these materials. The fact that these
films prepared at low thermal budgets have optimal optical absorption is a clear indication
of the resilience of the electronic properties against structural defects and 
vacancies~\cite{Kim2014}.
Experiments provide  few specific indications about point defects and intra-gap electronic levels. 
Electronic Paramagnetic Resonance (EPR) measurements detect CH$_2$NH$_3^+$ cations and Pb$^0$ 
paramagnetic clusters in samples which had been  irradiated with intense ultraviolet 
light~\cite{testedicazzo}.  
Evidence  of Pb metallic clusters in $\mathrm{MPbI_3}$ is also given  by X-ray
Photo-emission Spectroscopy~\cite{Leijtens2014,Conings2014};  ref.~\onlinecite{Leijtens2014} 
reports also the  presence of occupied (mainly in $n$-type conditions)  
or  empty (mainly in $p$-type conditions)   intra-gap electronic levels distributed at variable energies 
inside the gap. 
Indirect information about intra-gap electronic levels can also be drawn from
photoluminescence (PL) experiments. As an example, a recent work~\cite{Saba} revealed 
that while transient PL spectroscopy in the high excitation density regime is dominated by unbound
electron-hole radiative recombination, the steady-state PL intensity follows a 3/2-power
law as a function of laser intensity ($\mathrm{PL} \propto \mathrm{I}^{3/2}$) which deviates 
from a purely radiative behavior ($\mathrm{PL}\propto \mathrm{I}$) and suggests the presence of 
intra-gap states. These intra-gap states can trap either electrons or holes, but not both
of them simultaneously, thus indicating that these traps should be either fully occupied or
empty, but not partially filled. The trapping contribution is visible up to a injected carrier 
density of $\sim 10^{16}\div10^{17} \mathrm{cm}^{-3}$, above which the radiative regime takes place 
and the linear behavior $\mathrm{PL} \propto \mathrm{I}$  is recovered. First-principles
calculations are an extremely useful approach to evaluate the presence of defects in the
systems. In several recent theoretical works~\cite{Yin2014a,Buin2014,Duan2014} 
(see also Refs.~\onlinecite{Yin2014b,Yin2014} for a review) the
most common point defects (vacancies, interstitials, substitutionals, and anti-sites) were
thoroughly analyzed. It was found that the defects with the lowest formation energies (Pb
vacancy and interstitial methylammonium, according to Ref.~\onlinecite{Marchioro2014}) 
are shallow acceptors and donors, respectively, thus coherently with the expectation of 
good $p$-type or $n$-type conductivity and recombination properties dominated by radiative processes. 
For some defects with larger formation energies (e.g. Pb interstitial and 
anti-site~\cite{Marchioro2014,Buin2014,Du2014}) it is
estimated by the very simplified transition-state argument that electronic levels lying
well into the band gap could occur, but no explicit evidence of these defects was obtained
from the calculated electronic properties. In the search for localized states evidenced by
PL, in this work we use Density Functional Theory calculations to analyze a different
scenario, {\em i.e.} the possibility of intra-gap electronic localization induced by the presence
of interstitial molecules containing a C or N atom with a $2p$ lone-pair. 
We will consider only molecules that derive from the methylammonium dissociation or deprotonation. 
In doing so, we assume the rather conservative viewpoint of including only moieties that 
respect the chemical composition and stoichiometry of the bulk perovskite. Apart
from the dissociation processes,  these and other physically similar impurities can be 
introduced in $\mathrm{MPbI_3}$ accidentally. 
Nicely, this limited set of cases is
however sufficient to display a significant variety of defect states, whose characteristics
are consistent with the PL analysis.  In particular, we found that amine groups originating
from the lone-pair of nitrogen molecules (e.g. ammonia and methylamine produced by M
dissociation) have the general tendency to furnish trap states located well into the band gap. 

Fragments containing a carbon $2p$ lone-pair instead  are generally unstable and tend 
to form bonded complexes with I ions. In this case all  the localized electronic states are well below the valence band. 
This behavior is significantly influenced by the Fermi level position.  For example in $p$-type conditions 
the  CH$_2$NH$_3$  fragment yields one electron to (captures one hole from) 
the valence band and loosens its bond with iodine. The hole remains thus localized  on a  C($2p$) state of the fragment. 
This intragap  half filled level has been actually observed by EPR~\cite{testedicazzo}.  
    
Considering these defected perovskites as starting point, we then include interstitial Cl
doping in the analysis (previous calculations~\cite{Colella2013} only considered Cl as 
I-substitutional in the bulk perovskite, and as such, hardly capable to change 
on any significant extent the photo-absorption properties of the perovskite). 
Here we give evidence that additional inclusion of 2\% interstitial Cl in the perovskite 
(thus corresponding to $p$-type conditions) is capable to remove a corresponding 
fraction of N-derived localized states from the gap
without including any additional charge localization in its own. Our results provide a
clear evidence of the beneficial action of Cl in removing electronic localized states from the gap,
and furnish a possible explanation to the increase of an order of magnitude in lifetime and
diffusion length observed after Cl inclusion in solution. 

\section{Computational Methods}
The electronic and structural properties of the defects were computed using the  Density
Functional Theory within Local Density Approximation, as implemented in the ESPRESSO
package~\cite{QE-short}  which  is a well known plane-wave plus ultrasoft 
pseudopotential~\cite{Vanderbilt1990} code. 
Pseudopotentials for I, C and N have projectors on $s$  and $p$ channels. For C and N the
valence reference states are $2s$ and $2p$, and $5s$ and $5p$ for Iodine. The Pb pseudopotentials
have projectors in the $s$, $p$ and $d$ channels; with reference valence states $6s$, $6p$
and $5d$.

All calculations were done by using a plane-waves basis set cut-off  of  $35$~Ry. All the
reported results are inherent to the same $2\sqrt{2}\times 2\sqrt{2}\times 2$  supercell 
({\it i.e.}  16 MPbI3 cubic units),
which is the  $2\times 2\times 1$   repetition of the Tetragonal P4/mbm~\cite{Baikie2013}  
unit cell. 
Periodic boundary conditions are imposed along a tetragonal body centered lattice with 
vectors $(a, -a, c), (a, a, c), (-a, -a, c)$ , where $a = 8.712~\mathring{\text A}$  
and $c/a=1.41$  are the lattice parameters obtained by the theoretical optimization of 
bulk unit cell and in good agreement with those reported in Ref.~\onlinecite{Baikie2013} 
for the tetragonal phase which is most stable at room temperature.    

An accurate treatment of Pb $6p$ states in the conduction band would require the inclusion of 
Spin-Orbit Coupling~\cite{Even2013} (SOC). SOC has  nonetheless small
influence on energetics, structure and electronic properties of  valence band and  defects 
and  represents a major increase of the computational load. 
For this reason, similarly to what done in many other works on
this subject~\cite{Yin2014a,Buin2014,Duan2014,Marchioro2014} we choose to discard SOC. 
One additional advantage deriving from this choice is the quite realistic value for 
band gap yielded by LDA (albeit as a consequence of a fortuitous error cancelation).  

The integrations in the Brillouin Zone during self consistency uses a $3\times 3\times 3$ 
Monkorst-Pack mesh.  For the density of states calculation we have used the linear 
tetrahedron method on a 
$8\times 8 \times 8$  $k$-space mesh.

\section{Results and discussion} 
\subsection{Bulk perovskites} 
\begin{figure}
\includegraphics[width=\figwidth ]{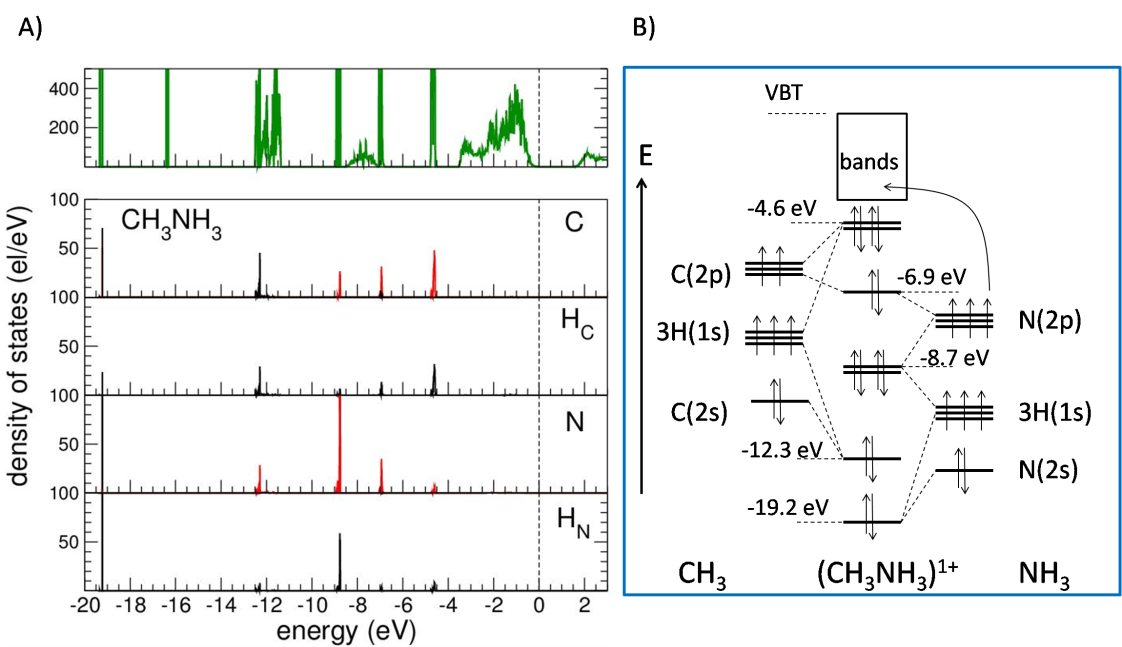} 
\caption{({\em Color on line.})
{\bf A}: Density of states of bulk MPbI$_3$, Valence Band Top is set to 0. 
To better visualize the molecular levels, they are
plotted in separate panels, one for each atomic type. Top panel: total DOS (green). Bottom
panels: projected DOS on atomic states of methylammonium molecule. 
HC and HN are hydrogens bound to C and N, respectively; 
black and red lines refer to s-type and p-type states. {
\bf B}: scheme of methylammonium orbital levels, as inferred for the calculated DOS.
\label{figura_1}
}
\end{figure}

In defect-free bulk perovskites, the M$^{1+}$ electronic states
are well localized in space, as seen from the calculated DOS in Figure~\ref{figura_1}. 

The electronic structure of methylammonium embedded in the perovskite is not dissimilar to
that of the isolated molecule, showing (from bottom, relatively to the Valence Band
Top(VBT)) a singlet (at $-19.2$~eV) derived
from N($2s$)-H($1s$) hybridized states; another singlet at $-12.4$~eV from C($2s$)-H($1s$)
hybridization; a N($2p$)-H($1s$) doublet at $-8.8$~eV, and another singlet at about $-7$~eV
derived from N($2p$)-C($2p$) states. Finally, the C($2p$)-H($1s$) doublet at $-4.8$~eV is 
the HOMO. Since this states are located well below the VBT, no molecular level interferes directly 
with the band gap region. The same holds for the empty molecular states (not shown here) being
the LUMO well above the Conduction Band Bottom (CBB).
This decoupling of extended bands (derived from Pb and I states) with respect to the molecular
states does not necessarily holds anymore if we consider configurations with methylammonium
broken in C-based or N-based fragments. In the following, we will examine all the possible
stoichiometric processes obtained by the fragmentation of methylammonium, giving rise to
stable molecules.
\subsection{Methylammonium fragmentation} 
A neutral M (i.e. not the M$^{1+}$ state of the bulk perovskite) lacks one H to be dissociated
in a couple of $\mathrm{NH_3}$ and $\mathrm{CH_4}$ molecules, which would be both stable and neutral. 
It follows that, forcing a starting configuration with two separated C- and N-based complexes, one of
them (the one lacking an hydrogen) will tend to form covalent bonding with neighboring
ions, while the most stable will remain located nearby the perovskite A-site. For
completeness we also include in the analysis the deprotonation processes.   Five  types of
decomposition leaving at least one stable product can be hypothesized:
{\bf a}) $\mathrm{CH}_3\mathrm{NH}_3 \to \mathrm{CH}_3 + \mathrm{NH}_3$;
{\bf b}) $\mathrm{CH}_3\mathrm{NH}_3 \to  \mathrm{CH}_4 + \mathrm{NH}_2$;  
{\bf c}) $2\times \mathrm{CH}_3\mathrm{NH}_3 \to \mathrm{CH}_3\mathrm{NH}_2 + 
           \mathrm{NH}_4+\mathrm{CH}_3$; 
{\bf d}) $\mathrm{CH}_3\mathrm{NH}_3 \to  \mathrm{CH_3NH_2} + H$; 
{\bf e}) $\mathrm{CH_3NH_3} \to \mathrm{CH_2NH_3}+\mathrm{H}$.

These five processes conserve the perovskite bulk stoichiometry so that it is possible to
estimate their energetic cost by direct comparison of total energies.  The results for the
completely separated fragments are reported in Table~\ref{table_1}. 
The computations show that the $\mathrm{CH_3\tract I}$ fragment has a significant binding energy with 
$\mathrm{NH_3}$ ($0.4$~eV) and $\mathrm{CH_3NH_2}$ ($0.5$~eV).

We should notice that our structures are product of atomic relaxations which sample a 
limited portion of the potential energy profile. Thus, we cannot exclude that the same 
final products could be arranged in a more energetically convenient way than that found 
in our final state.

\begin{table}
\caption{
Energetic cost of the fragmentation processes($\mathrm{\Delta E}$) 
examined in this work. The energies are calculated with respect to an 
equivalent bulk $\mathrm{MPbI_3}$ supercell with 16 cubic perovskite units. The
corresponding Average bond dissociation  enthalpies ($\mathrm{\Delta H}$) 
are reported, as a comparison  with the corresponding dissociation.
\label{table_1}
}
\begin{ruledtabular}
\begin{tabular}{cdd}
Configuration&
\multicolumn{1}{c}{$\mathrm{\Delta E}$ (eV)}
&
\multicolumn{1}{c}{$\mathrm{\Delta H}$ (eV)}
\\
\hline
$\mathrm{NH_3}+\mathrm{CH_3}\tract\mathrm{I}$& 2.03& 3.17\\
$\mathrm{CH_4}+\mathrm{NH_2}\tract\mathrm{I}$& 1.90& 3.17\\
$\mathrm{CH_3NH_2I}+\mathrm{NH_4}+\mathrm{CH_3}\tract\mathrm{I}$& 2.30 & 3.17\\
$\mathrm{CH_3NH_2I}+\mathrm{H}\tract\mathrm{I}$& 2.66& 4.03\\
$\mathrm{CH_2NH_3}\tract\mathrm{I}+\mathrm{H}\tract\mathrm{I}$& 3.22&4.28\\
\end{tabular}
\end{ruledtabular}
\end{table}

As one can see in Table~\ref{table_1} the formation energies $\mathrm{\Delta E}$  substantially 
reflect the average bond enthalpy $\mathrm{\Delta H}$  of the C-N bond in case of {\bf a}) and {\bf b}) 
configurations (see the Table), and of the N-H bond in case of {\bf c}). 
Since the dissociation products are stable molecules, they gain some stability at the end of
the process, thus $\mathrm{\Delta E}$  are actually sizably smaller than $\mathrm{\Delta H}$. 
These large formation energies are indeed
coherent with the low concentration of localized defects indicated by the experiments. However, this
should not lead to the conclusion that the final products of dissociation are unlikely, since they
could form during non-equilibrium solution processing, and not necessarily through the actual
thermodynamic dissociation of methylammonium. Further, as reported in Ref.~\onlinecite{testedicazzo}
deprotonated fragments are  formed upon ultraviolet irradiation and, once created, 
they are stable. 
Finally notice that these results should not be compared 
with those calculated  for point defects in  Refs.~\onlinecite{Yin2014b,Marchioro2014,Roiati2014}: 
in that case \emph{non-stoichiometric} defects are considered 
(e.g. vacancies, interstitials, substitutionals) thus a suited choice of atomic chemical 
potentials can always favor the formation of  specific defects. In the present
case, all the configurations are stoichiometric, and no chemical reservoir is assumed. 
The resulting energies depend only on the actual  energetic cost of the dissociation mechanism
and do not depend on adjustable parameters. 
In the following we will analyze structural and electronic properties of the fragments. 

\subsection{Dissociation of methylammonium in ammonia and iodomethane}
\begin{figure}
\includegraphics[width=\figwidth ]{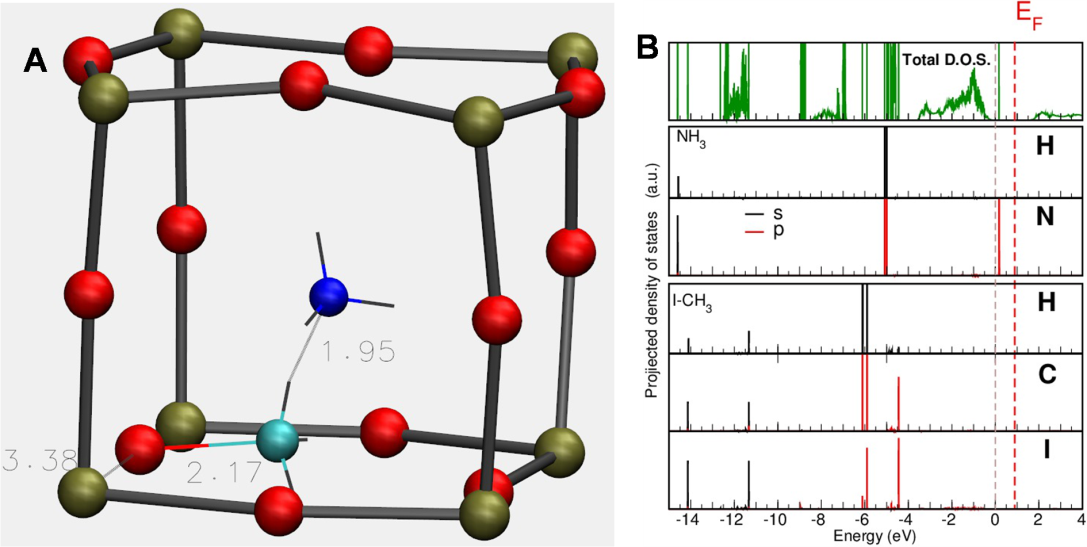}
\caption{({\em Color on line})
Atomic and electronic structure of the perovskite after $\mathrm{CH_3NH_3}\to \mathrm{CH_3} +
\mathrm{NH_3}$  a) dissociation process. 
{\bf A}): Atomic structure in close-by configuration: $\mathrm{CH_3}$ migrates onto a 
$\mathrm{PbI_4}$ plaquette to form a iodomethane molecule ($\mathrm{CH_3I}$, 
the C−I dimer indicated by cyan and red balls), while $\mathrm{NH_3}$ remains close
to the initial A-site of the perovskite. 
{\bf B}): corresponding DOS. Upper panel: total DOS (green);
lower panels: atom-resolved DOS for $\mathrm{CH_3I}$ and $\mathrm{NH_3}$
 molecules (orbital contribution labels are the same as the DOS in Figure~\ref{figura_1}). 
Zero energy is fixed at the VBT.
\label{figura_2}
}
\end{figure}

Following the mechanism (a) we  forced the separation of one M in $\mathrm{NH_3}$ (ammonia) 
and $\mathrm{CH_3}$. 
The former remains close to the starting M position (i.e. the A-site of the perovskite) and fairly 
stable in its ground state, with three electrons distributed in an equal number of N-H bonds, 
and other two in a lone-pair HOMO of N($2p$) orbital character. On the other hand, $\mathrm{CH_3}$ ­
is in a highly electronegative state, it migrates toward an I atom 
(also in the need of an electron to fill the valence hole, assuming a neutral M at the start) 
After structural optimization, a stable $\mathrm{CH_3}\,\tract\,\mathrm{I}$ complex 
(\emph{"methyl-iodide"}, also called \emph{"iodomethane"})
is established, with the C-I axis lying nearly parallel to the $\mathrm{PbI_4}$ square and a
bond length of $2.17~\angst$. The key feature, 
for what concern the photo-absorption properties, is that the N($2p$) HOMO level of $\mathrm{NH_3}$  
is high enough in energy to overcome the whole I($5p$)-Pb($6s$) valence 
band manifold (spanning a $3.6$~eV wide energy interval) and intrudes well within the band gap. 
Thus, a doubly occupied donor appears $0.17$~eV above the VBT.  
For iodo-methane, on the other hand, the HOMO state is a highly stable I($5p$)-C($2p$)
$\mathrm{\sigma}$-bond lying well below the valence band manifold, thus ineffective 
for what concerns the band gap region.
The detailed properties of this configuration are shown in Figure~\ref{figura_2}.

\subsection{Dissociation of methylammonium in methane and iodoamine}
\begin{figure}
\includegraphics[width=\figwidth ]{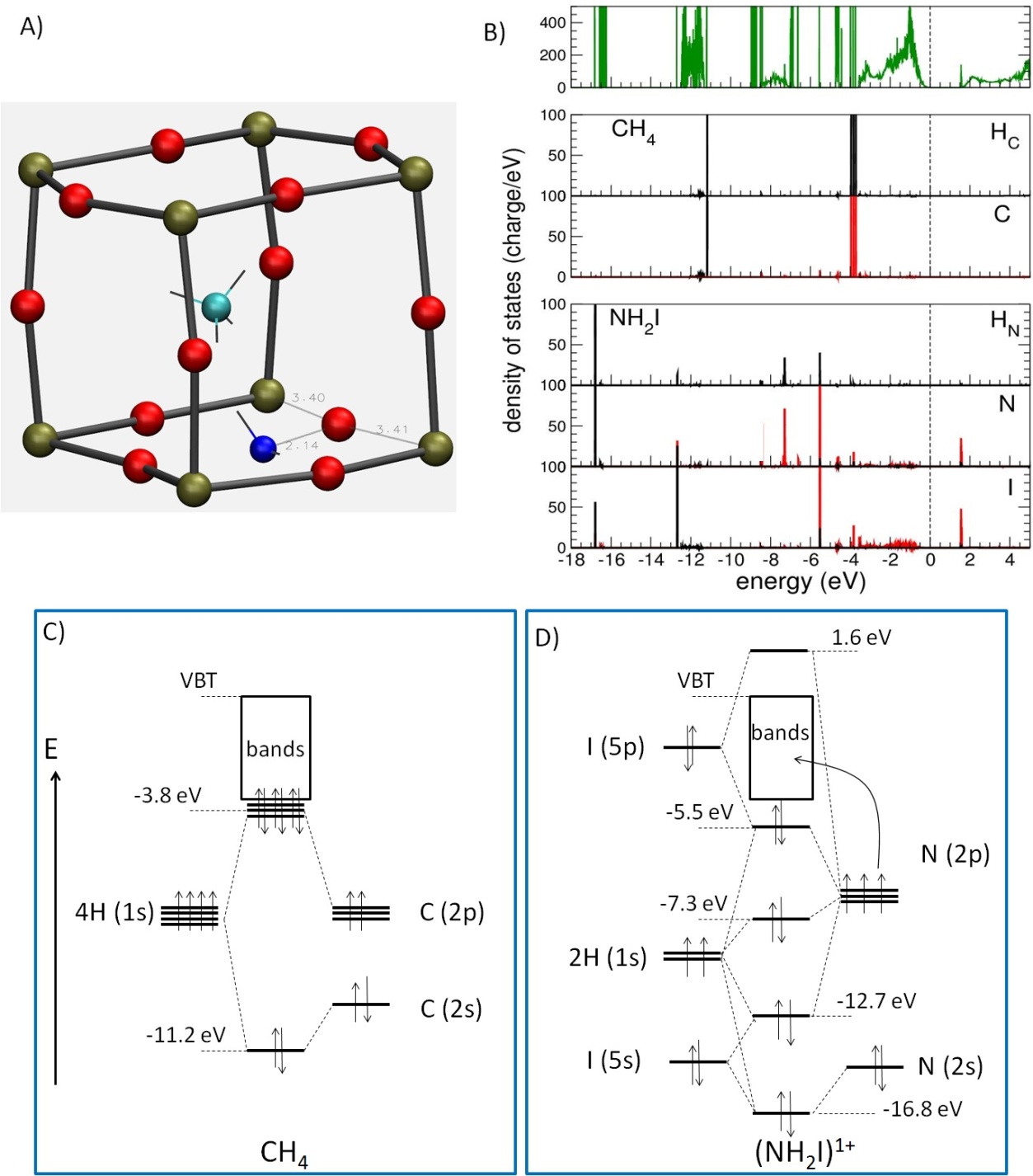}
\caption{({\em Color on line.})
Structural optimization of the $\mathrm{CH_3NH_3}\to \mathrm{CH_4} + \mathrm{NH_2}$ 
dissociation c); {\bf A}) $\mathrm{NH_2}$ migrates into a $\mathrm{PbI_4}$
plaquette to form a iodoamine like complex ($\mathrm{NH_2}\tract\mathrm{I}$), and methane 
($\mathrm{CH_4}$) remains stuck to the starting M-site. 
{\bf B}): corresponding DOS, and atomic projected DOS. 
Orbital types and labels are the same as in Figure~\ref{figura_2}.
{\bf C} and {\bf D}): schemes of the molecular levels, as inferred from the atomic-projected
DOS.
\label{figura_3}
}
\end{figure}

Following mechanism (b) we separate methylammonium in $\mathrm{CH_4}$ (methane) and $\mathrm{NH_2}$, 
the behavior of C and N is exchanged with
respect to the previous case: the stable methane molecule remains nearby the starting M site, while
$\mathrm{NH_2}$ migrates toward a I ion  and forms a $\mathrm{NH_2I}$ (iodoamine) 
molecule again placed at the interstice  of the perovskite's octahedral cage. The electronic structure of $\mathrm{CH_4}$ (shown in Figure~\ref{figura_3}) 
is quite close to that of an isolated methane molecule, indicating 
small interaction with the inorganic surroundings. However,
the electronic levels of this fragment remain far from the important energy region of
photoabsorption. The interstitial $\mathrm{NH_2}−\mathrm{I}$ 
fragment also shows an electronic structure substantially
similar to that of the isolated iodo-amine molecule, with the LUMO given by a 
N($2p$)-I($5p$) hybridized
state lying just at the bottom of the conduction band. Thus, this state is a deep acceptor, but with
a binding energy too small to be effective as recombination center.

\subsection{Formation of ammonium, iodomethane and methylamine.}
Following dissociation process (c), specularly to path (b), we separate the methylammonium in $\mathrm{NH_4}$ and $\mathrm{CH_2}$ 
fragments. Again the stable $\mathrm{NH_4}$ remains at the A-site and the unstable fragment is attracted by one I ion. 
In this case the $\mathrm{CH_2}$  not only  binds to  a I ion but --as it has to stabilize two $2p$ one pairs--  
captures also one  H from a close-by methylammonium (thus ending up to a methylamine) and  forms  
a stable $\mathrm{CH_3} \tract \mathrm{I}$ {\em iodo-methane} complex. 
This result confirms that the formation of $2p$ lone pair on N is energetically more favorable with respect to C. 

  Structural and electronic properties of methylamine and iodo-metane are illustrated in the discussion of dissociation 
paths (b) and (d), respectively. The 4 electronic levels stemming from ammonium valence states are all well below the I($5p$) valence 
band.  1 level at about  $-20$~eV and 3 almost degenerate levels at about $-10$~eV with respect to the VBT.   

\subsection{Dissociation of methylammonium in hydrogen iodide and methylamine.}
\begin{figure}
\includegraphics[width=\figwidth ]{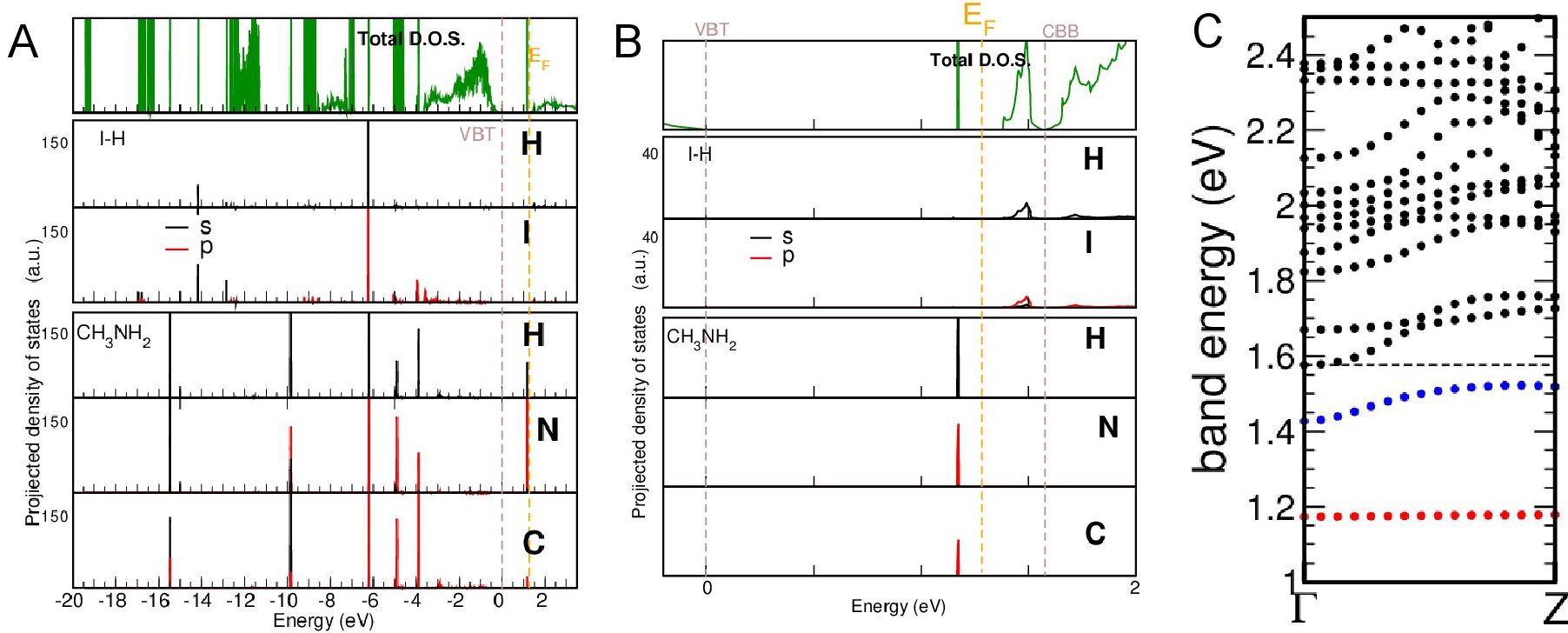}
\caption{({\em Color on line.})
Simulation of the $\mathrm{CH_3NH_3} \to \mathrm{CH_3NH_2} + \mathrm{H}\,\tract\,\mathrm{I}$ 
dissociation: one H migrates into a $\mathrm{PbI_6}$ octahedron to form hydrogen iodide (HI), 
leaving a stable methylamine molecule ($\mathrm{CH_3NH_2}$) close to the original M-site.
{\bf A}):  atomic resolved DOS in the full energy range.
Orbital types and labels are the same as in Figure~\ref{figura_1}-A. {\bf B}) DOS in a small energy 
interval around the band gap, to better visualize the molecular levels overlapping 
with the valence bands (see text). {\bf C}): band structure alog the $\Gamma\tract Z$ direction. 
The flat red band is due to the N($2p$) lone-pair levels of methylamine. The blue band is related to H-I empty levels.
The dashed line indicates the level of the Conduction Band Bottom.
\label{figura_4}
}
\end{figure}

The methilamine can also form directly from the deprotonation process of path (d).  
 The detached H  migrates toward  the $\mathrm{PbI_6}$ octahedron and binds to a I to form a HI molecule (hydrogen iodide).
The deprotonated  $\mathrm{CH_3NH_2}$ (methylamine) molecule is stable (structural and
electronic properties are shown Figure~\ref{figura_4}) and thus remains in a substitutional position at the A-site. 
At variance with M, methylamine is neutral, thus the I($5p$) valence band manifold remains nominally hole-doped. 
Our calculations show that this hole is compensated by the HI formation, which produces localized $s\tract p$  states, thus
formally subtracting one I to the Pb-I bands. But the most interesting aspect is that the
filled methylamine HOMO and the empty hydrogen iodide LUMO are both located well within the band gap.

For what concerns methylamine, its 14 electrons fill 7 doubly-occupied localized levels.
The second-highest in energy (a C-H bond at $-2.9$~eV) falls well within the I($5p$) valence 
manifold an it is slightly broadened and barely visible on the same scale of the other levels 
of the molecule ({\em see the DOS enlargement in a small energy region around the band gap in
Figure~\ref{figura_4}-panel {\bf C}}). 
The highest occupied state of methylamine is a N($2p$) lone-pair, located $1.17$~eV above the VBT; 
thus, as in the case of the the ammonia molecule, we have a stable molecule whose HOMO, derived by a
lone-pair of N($2p$) character, is high enough in energy to emerge above the VBT. For what concerns
the HI molecule, it presents a I($5p$)-H($1s$) singlet at $-6.2$~eV, and a I($5p$)-H($1s$) doublet at 
$-3.9$~eV.
Interestingly, its first empty state is another I($5p$)-H($1s$) bond below the CBB. To highlight
the characteristics of these in-gap states we also calculated the band dispersion along
the $\mathrm{\Gamma-Z}$ direction (Fig.~\ref{figura_4}--D): while the $\mathrm{CH_3NH_2}$ 
HOMO is quite flat through the whole region, the HI LUMO is characterized
by a substantial broadening ($\approx 0.1$~eV), remaining however separated from the CBB 
by $\sim\!\!50$~meV. 
The presence of these fully occupied and empty in-gap states is a potentially relevant feature 
for what concern photo-absorption and photo-emission, since they represent sources of 
strong non-radiative electron-hole recombination processes.

\subsection{ Dissociation of methyl-ammonium in hydrogen iodide and $\mathrm{CH_2NH_3\tract I}$}
\begin{figure}
\includegraphics[width=\figwidth ]{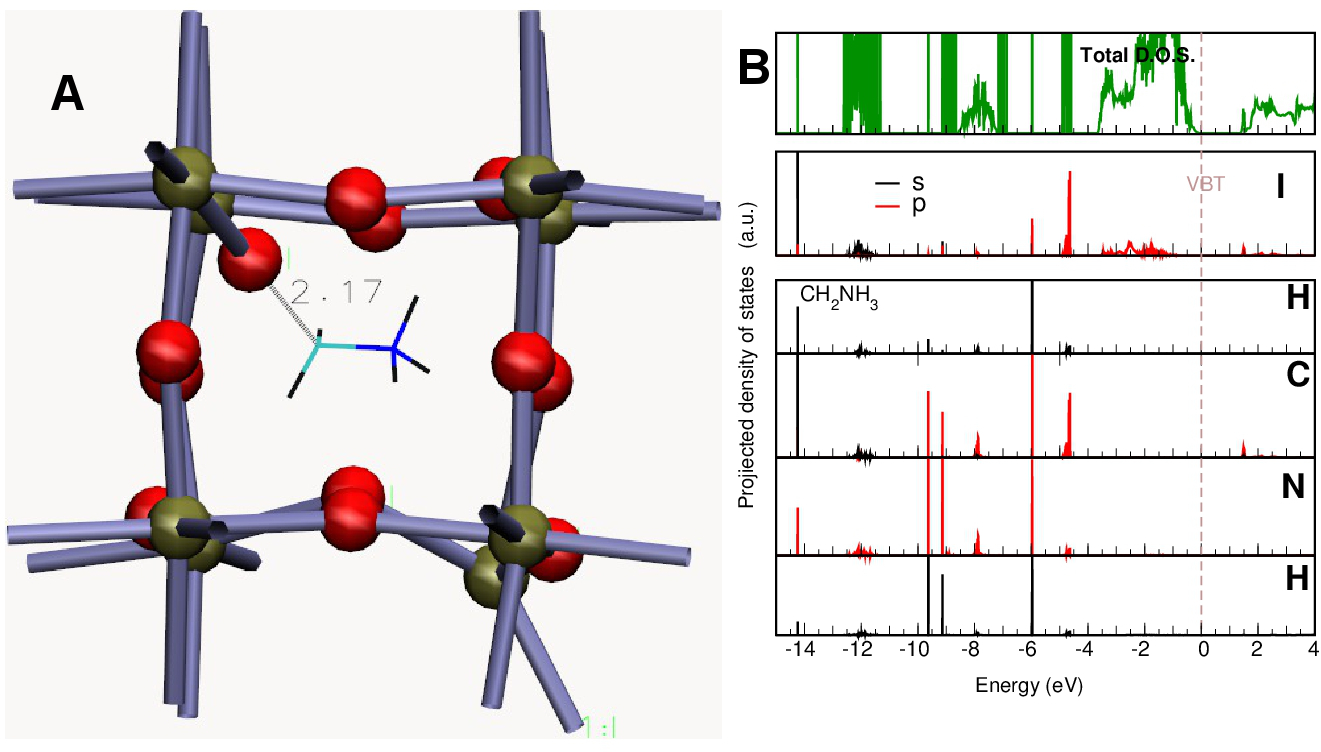}
\caption{({\em Color on line.})
{\bf A}) Structure of the $\mathrm{CH_2NH_3} \tract \mathrm{I}$ complex. 
The I ion is displaced from its equilibrium position to bind to the deprotonated fragment. 
{\bf B}) electronic structure of the  $\mathrm{CH_2NH_3} \tract \mathrm{I}$  complex, 
illustrated by the DOS projected on $\mathrm{CH_2NH_3}$ 
and I sites. States  are filled up to  the Valence Band Minimum (VBM).
Except for the H-I peak close to the conduction band there are no other defect states within the
gap. Five distinct peaks ({\em see text} ) are present in the projected DOS of $\mathrm{CH_2NH_3}$; 
C-derived $s \tract p$ states are strongly hybridized with the I $5p$  states.
\label{figura_5}
}
\end{figure}

Following process (e) the  deprotonation occurs at the C side.  
We have again the formation of an H-I  dimer which implies the transfer
of one electron to the I($5p$) valence band. The deprotonated methyl-ammonium fragment is thus
neutral.  At variance with the $\mathrm{CH_3NH_2}$ fragment,  $\mathrm{CH_2NH_3}$ binds to
an I  ion and forms  a stable complex which is in many aspects analogous to the
$\mathrm{CH_3\tract I}$ complex seen above.  Also the  computed bond length ($2.17~\angst$)  
is the same as we obtained for the iodo-methane complex. 
All of the 14 electrons of this fragment are localized in molecular levels whose energies are
well below the valence band~({\em see Figure~\ref{figura_5}}). With respect to the valence
band top (VBT) we have 3 distinct levels at  $-14$~eV, $-9.5$~eV and $-6$~eV; we have 4 more 
electronic levels that are organized in two doublets at  $-9$~eV and at $-4.8$~eV. The
electronic orbitals relative to the doublet at $-4.8$~eV are the bond orbitals of the
complex showing a strong hybridization between I($5p$) and C($2p$) atomic orbitals.     

\subsection{Effects of the $p$-type doping.}
In the stoichiometric dissociation processes described above the loss of a neutral M leaves the
I($5p$) valence band manifold with one electronic vacancy. This hole is compensated when a 
dissociation fragment binds to a I atom (forming $\mathrm{H \tract I}$ ,
$\mathrm{CH_3 \tract I}$, $\mathrm{NH_2 \tract I}$, or
$\mathrm{CH_2NH_3 \tract I}$  complexes) so that no
holes are present at the VBT. Starting from these stoichiometric configurations,
 in order to explore the effect of $p$-doping on the intra-gap localized states, 
we further enforced $p$-type conditions by subtracting electron charges from the supercell.  
Surprisingly, in the case of methylamine or ammonia, the filled lone pair is not ionized by 
$p$-doping, instead it  shifts down in energy below the Fermi level and out of the band gap.  
If, by an increase of hole concentration, we shift further down the Fermi level, the lone-pair
remains below the Fermi level undergoing rigidly to the same energy shift. 
An outstanding consequence then results from our simulations: amine lone-pair levels can stay in the
gap as long as they are fully occupied or empty, but not if partially occupied, since in this case
they regain full occupancy by shifting in energy below the VBT and thus taking  one electron charge 
from the I($5p$) valence bands. According to these  results  thus it is impossible to have N($2p$) lone-pair 
levels with partial filling.  As previously mentioned, the lack of 
in-gap partially filled states is in agreement with the arguments based on PL experiments of Refs.~\onlinecite{Saba} 
and~\onlinecite{Leijtens2014}. 
A different situation is obtained by $p$-doping path (e) leading to the formation of the
$\mathrm{CH_2NH_3}\,\tract \,\mathrm{I}$ complex: 
the HOMO doublet of this complex,  which in intrinsic conditions is a strongly covalent
superposition of C($2p$) and I($5p$) orbitals, in consequence of  $p$-doping splits 
in a fully occupied C($2p$) singlet --laying about $5$~eV below the VBT --and in a half-filled 
localized state inside the band gap-- at about $1.4$~eV above the VBT.  
This result is coherent with EPR measurements in irradiated $\mathrm{MPbI_3}$~\cite{testedicazzo}
which indicate the presence of half-filled electronic orbitals derived by $\mathrm{CH_2NH_3}^+$ 
ions.

\begin{figure} 
\includegraphics[width=\figwidth ]{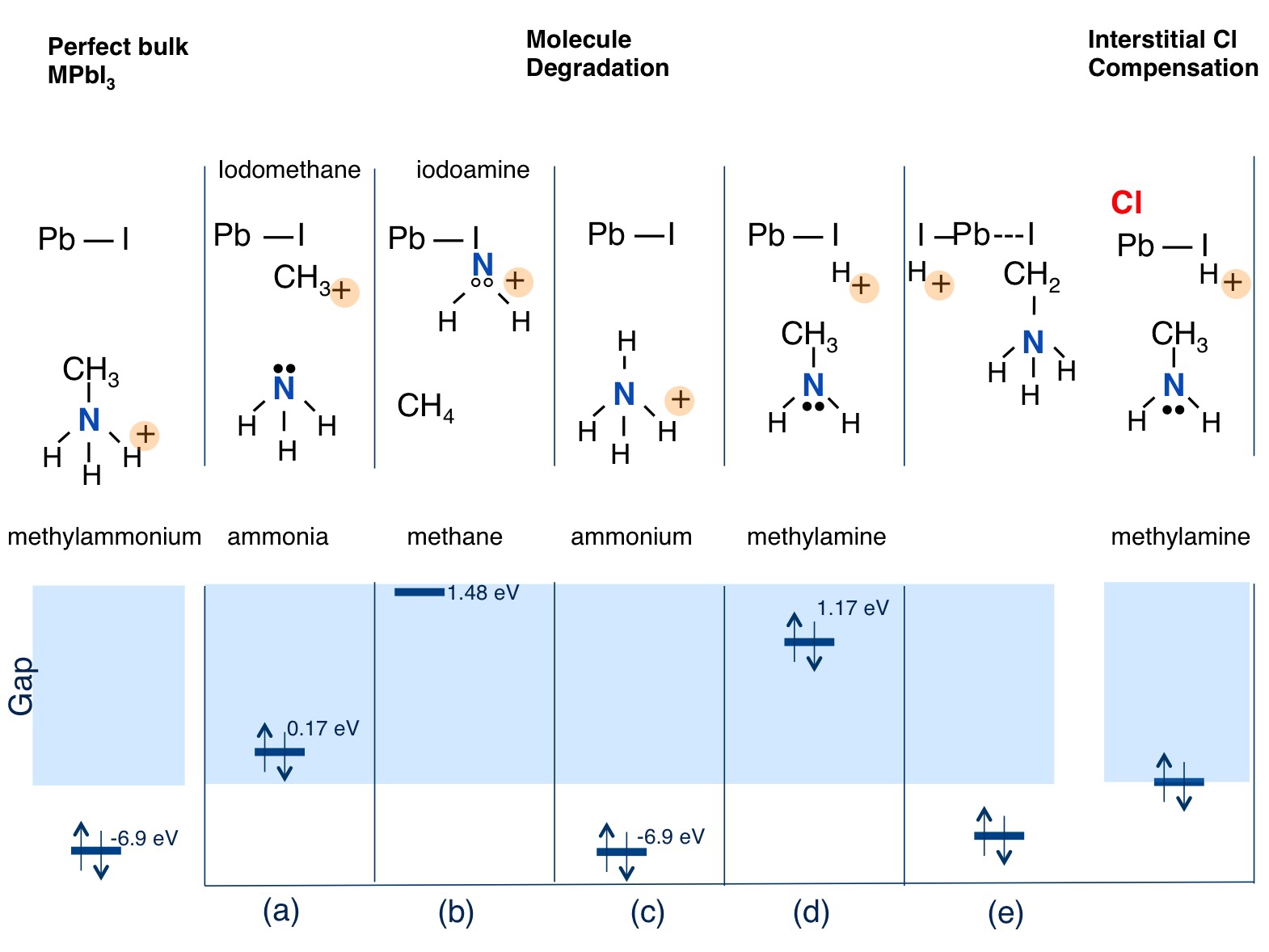}
\caption{({\em Color on line.})
Scheme of the stable fragments considered in the present work under stoichiometric conditions. For
each configuration the molecular levels closer to the band gap region are also indicated. In
particular, amine groups display localized lone-pair HOMO and LUMO orbitals well into the band gap
which are eventually removed after addition of interstitial Cl.
\label{figura_9}
}
\end{figure}

\begin{figure}
\includegraphics[width=\figwidth ]{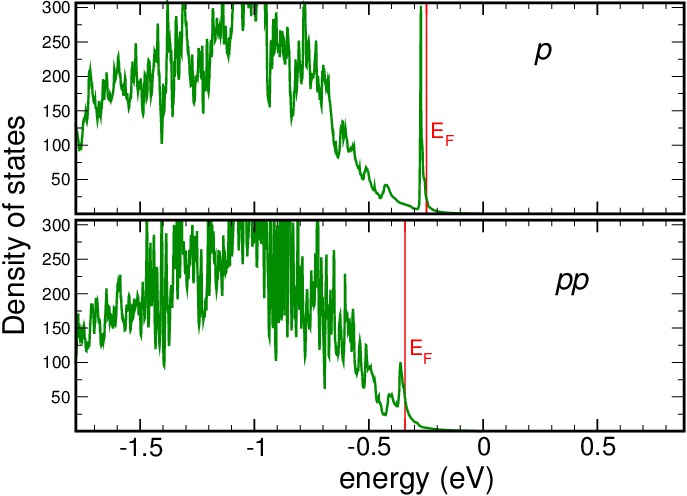}
\caption{({\em Color on line.})
Density of states close to the VBT for two $\mathrm{MPbI_3}$  supercells containing one 
methylamine at different hole doping concentrations. Top-panel: a bulk $\mathrm{MPbI_3}$ 
with one methylammonium substituted by one methylamine (i.e. one H subtracted) in the 192-atom supercell. 
This corresponds to a hole doping concentration of $2.74\times10^{20} \mathrm{cm}^{-3}$. 
Bottom panel: a bulk $\mathrm{MPbI_3}$ plus a M-vacancy plus a M-to-methylamine substitution 
(2 holes per supercell).
\label{figura_6}
}
\end{figure}

In summary, according to our results N-derived and C-derived localized in-gap states behaves quite
differently: the former can never stay partially occupied in the band gap, and subtract electrons from
the valence band if $p$-doping conditions are enforced; the latter, on the other hand remains well
within the band gap even if half-occupied. We can understand this difference on the basis of the
very high electronegativity of $p$-doped N-H group. In comparison, the electronegativity of C-H
molecules is much lower and these latter  have thus the tendency to form hybridized bonds with I ions or alternatively 
strip H from neighboring M molecules.  
Finally, the
total energy comparison between perovskites including methylamine and $\mathrm{CH_2NH_3}^+$ 
in the same $p$-type conditions (thus directly comparable having the same number of atoms in the 
simulation supercell) shows that the latter is $0.46$~eV/f.u. higher in energy. 
Thus, the dissociation of methylammonium in methylamine should be regarded as the most probable 
source of in-gap localized defects in the perovskite.

\begin{figure}
\includegraphics[width=\figwidth ]{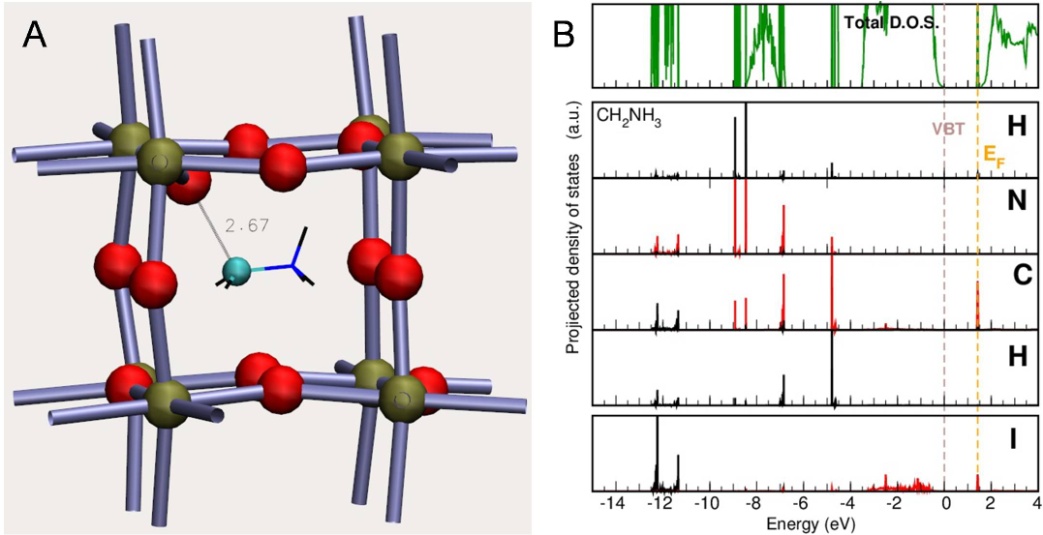}
\caption{({\em Color on line.})
 {\bf A}) Atomic structure of the $\mathrm{CH_2NH_3}^+ \tract \mathrm{I}$ complex 
in $p$-type conditions. {\bf B}) Electronic structure of the complex illustrated by the DOS 
projected on the atomic orbitals of the molecules.
\label{figura_7}
}
\end{figure}

\subsection{Role of Cl in perovskites.}
\begin{figure}
\includegraphics[width=\figwidth ]{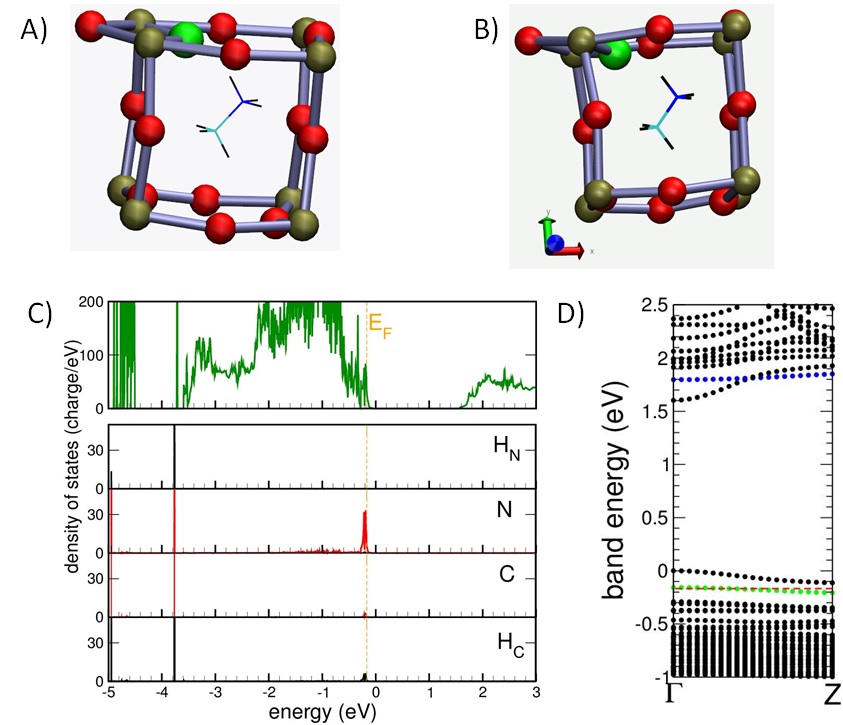}
\caption{({\em Color on line.})
{\bf A}) Atomic structure obtained upon structural optimization after inclusion of one Cl in a bulk-like
MPbI3 supercell; {\bf B}) atomic structure upon structural optimization with Cl included in a supercell
after M dissociation in methylamine plus hydrogen iodide. In both cases, Cl ends up in the
octahedral cage, binding with Pb atoms. C) DOS relative to the system displayed in B). The upmost
panel is the total DOS, lower panel the DOS of the methylamine molecule, resolved per atoms. As
usual red and black colors represent $p$ and $s$ states. {\bf D}): band structure for the same system along
$\mathrm{\Gamma}\tract\mathrm{Z}$ . The green line is the HOMO of methylamine, 
the blue line a HI state, both laying within the band gap before Cl inclusion (Fig.~\ref{figura_5}-D).
\label{figura_8}
}
\end{figure}

A major conundrum concerns the capability of Cl to enhance the photoconversion properties of the
perovskite; since the very first work revealing the solar cell efficiency of these
systems~\cite{Lee2012}, the
presence of Cl in solution with the reactants of the perovskite was highlighted as a key ingredient
to obtain power conversion efficiency (PCE) greater than 10\%. This is somewhat surprising since on
the one hand, the actual amount of Cl included in the perovskite was revealed to be not larger than
few percent and possibly accumulated at the interfaces~\cite{Colella2014,Mosconi2014}; 
on the other, it was realized that the isovalent Cl-I substitution could bring little effect on 
the fundamental electronic properties of the system~\cite{Colella2013}; in fact, 
several PL studies highlighted the fact that Cl could increase the lifetime
and the related diffusion length of the perovskite by an order of magnitude~\cite{Stranks2013}, 
a result hardly understandable in terms of a few-percent I$\,\to\,$Cl substitutions. 
These considerations have motivated a thorough search of the possible mechanisms which could explain 
the role of Cl in the perovskites. At the present, one of the most credited hypothesis focuses 
on the role of the interfaces where a significant concentration of Cl could be confined, 
thus effectively changing the photoabsorption properties~\cite{Colella2014,Mosconi2014}. 
An alternative explanation, very interesting in its simplicity, is based on the
hypothesis that a certain amount of native defects present in the undoped stoichiometric perovskite
could be removed from the gap  by the inclusion of interstitial Cl, acting as a  $p$-type
dopant. 
An estimate based on purely radiative band-to-band recombination~\cite{Filippetti2014}
showed that a reduction of defect concentration of about one order of magnitude can justify an
increase of lifetime of the order of that found in PL experiments. This scenario implies that: i)
native doping in the perovskite is $n$-type; ii) interstitial Cl doping is somehow capable to
eliminate the defect states laying in the band gap. In our description of methylammonium
dissociations, we found that condition i) is actually realized by two configurations characterized
by HOMO levels well into the band gap (see a summary of results in Figure~\ref{figura_9}). 
In the following we will show that even the hypothesis ii) is absolutely justified by our calculations, 
{\em  i.e.}  the inclusion of interstitial Cl always acts in a way to compensate the trap states and 
remove them from the band gap.  We have simulated the inclusion of a Cl atom in the perovskite,
together with the most interesting end-products of methylammonium, for various Cl starting
positions.  
The inclusion of one Cl in our 16 f.u. supercells corresponds to a 2\% additional concentration with
respect to the I atoms, in line with the PL estimates of defect concentration. Interestingly, we
found a rather universal behavior for the interstitial Cl, {\em i.e.} scarcely dependent on the initial
atomic configurations: in all the cases, Cl migrates into the octahedral cage, bonding with one Pb
and distorting the Pb-I octahedral cage; in Figure~\ref{figura_8} A) and B) we report the final 
structure for two different simulations, one with an additional Cl atom included in the bulk 
$\mathrm{MPbI_3}$ perovskite, and the other with a Cl included together with $\mathrm{CH_3NH_2}$ 
and HI as products of the methylammonium dissociation.
The two atomic structures in the region around Cl are quite similar, with Cl moving close to
a Pb and sharing the ligand position with one I ion. In this way Cl($3p$) states are
hybridized with the I($5p$) valence states, thus effectively
decreasing the $n$-type concentration by one electron per Cl. In Figure~\ref{figura_8} we report DOS 
and band energies relative to the relaxed structure with one $\mathrm{CH_3NH_2}$  plus HI plus Cl. 
The HOMO state of methylamine, which in absence of Cl was well into the band gap, is now located 
just below the VBT, {\em i.e.} the inclusion of Cl cleans up the band gap from its occupied localized 
states.

This {\em cleaning} mechanism is rather transparent: Cl grabs an electron from the in-gap lone-pair,
which then falls down in energy as a consequence of the subtracted electrostatic repulsion.
Furthermore, the $p$-type states of Cl spreads through the I($5p$) valence band region, without
introducing any relevant change in the total DOS, i.e the basic electronic properties of the
perovskite are not disrupted by the interstitial Cl, at least for the 2\% concentration examined
here. We emphasize that this mechanism is quite general: any $n$-type dopant, whatever the source,
will be compensated by an equal concentration of interstitial Cl subtracting one electron from the
valence band manifold.

\section{Conclusions}
In summary, we have studied by First-Principles calculations the effect in the perovskites of
various molecular defects. We have focused our analysis on those which can be obtained by the
dissociation or fragmentation of methylammonium. For these defects the formation energies calculated
by total energy differences range from about $2$~eV up to $3.2$~eV. Accordingly their
concentration is expected to be low under thermal equilibrium conditions, but it can eventually
increase out of equilibrium, i.e. during synthesis or as a result of chemical contamination.
Our results demonstrate that some of these defects form (either completely filled or empty)
localized levels inside the band gap, in particular those related to N($2p$) lone pairs present in
ammonia and methylamine molecules derived from methylammonium dissociation. An overall account of
these defects is schematically summarized in Figure~\ref{figura_9}: while the electronic states of
the $\mathrm{M}^{1+}$
molecule in the bulk perovskite lie far beyond or above the relevant band gap extremes, these
molecules display either HOMO or LUMO states well within the band gap (specifically, the ammonia
HOMO $0.17$~eV above the VBT, and the methylamine HOMO and LUMO $1.17$~eV and $1.48$~eV above VBT,
respectively). 

Also, we have shown that the amine lone-pair localized in the band gap are always
either full or empty, since any attempt to extract an electron from the in-gap HOMO state results in
a sudden falling of this state below the VBT. In other words, their fractionally occupied states are
never stable within the band gap. This result is in striking agreement with the interpretations of
the steady-state PL experiments carried out in Ref.~\onlinecite{Saba}.

\begin{figure}
\includegraphics[width=\figwidth ]{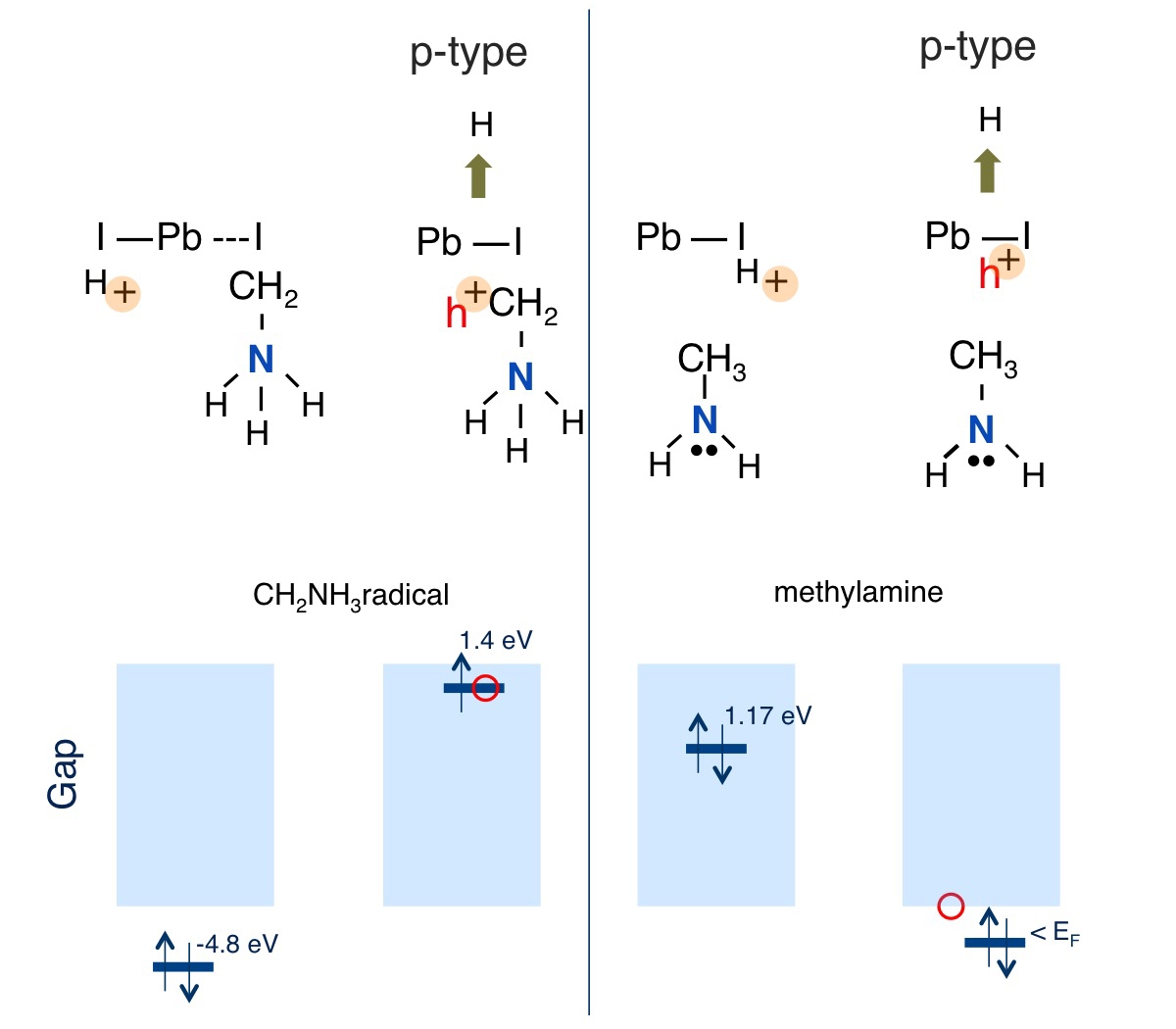}
\caption{({\em Color on line.})
Schematic representations of the effects of $p$-type doping; (left) the molecular level of the
methyl-type fragment is able to trap one hole and it forms a singly occupied level right in the gap;
(right) the lone pair of the methylamine fragment is always doubly occupied, and shifts below the
Fermi level in $p$-type conditions. 
\label{figura_10}
}
\end{figure}

At variance with amines, fragments having a C($2p$) lone-pair like $\mathrm{CH_2NH_3}$ or $\mathrm{CH}_3$ can form stable 
complexes by binding to I ions, but, as long as stoichiometric conditions are retained, 
their HOMO levels always remain well below the VBT.  However, in $p$-type conditions the orbital
related to C-I bond remains singly occupied thus moving into the band gap; this is in agreement with EPR
evidences, reported in Ref.~\onlinecite{testedicazzo}, of half-filled orbitals derived by 
the $\mathrm{CH_2NH_3}$  radical. Results for p-type conditions are summarized in Figure~\ref{figura_10}. 
Furthermore, we have analyzed the inclusion of interstitial Cl in the perovskite. Our results show
that an amount of 2\% interstitial Cl is capable to remove from  the band gap from an equal amount of
N-derived lone-pair HOMO states of methylamine or ammonia, and at the same time does not disrupt the
fundamental electronic properties of the perovskite. Our results build a sound conceptual fundament
to the observed enhancement of lifetime and diffusion length caused by the inclusion of Cl in the
solution processing of the perovskite.

\begin{acknowledgments}
The Authors acknowledge financial support by CompuNet, Istituto Italiano di Tecnologia (IIT), 
by Regione Autonoma della Sardegna under L. R. 7/2007 (CRP- 24978 and
CRP-18013), by Consiglio Nazionale delle Ricerche (Progetto Premialità RADIUS), by Fondazione Banco
di Sardegna Projects n. 5794 and n. 7454. They also acknowledge computational support by CINECA
(Casalecchio di Reno, Italy) and CRS4 (Loc. Piscina Manna, Pula, Italy).
\end{acknowledgments}

%

\end{document}